\documentclass[showpacs,prl,twocolumn,floatfix,reprint]{revtex4-2}
\usepackage{amsmath,graphicx,amssymb,hyperref}
\usepackage{mathtools, stmaryrd}

\usepackage[english]{babel}

\usepackage[letterpaper,top=2cm,bottom=2cm,left=3cm,right=3cm,marginparwidth=1.75cm]{geometry}

\usepackage{amsmath}
\usepackage{graphicx}
\usepackage{hyperref}
\usepackage{xcolor}

\usepackage{graphicx}
\usepackage[caption=false]{subfig}

\newcommand{\rev}[1]{{\color{black}{#1}}}

\begin{document}

\title{A Stochastic Block Hypergraph model}

\author{Alexis Pister}
\affiliation{University of Edinburgh, 1 Lauriston Pl, Edinburgh EH3 9EF, United Kingdom
}

\author{Marc Barthelemy}
\affiliation{Universit\'e Paris-Saclay, CNRS, CEA, Institut de Physique Th\'{e}orique, 91191, 
	Gif-sur-Yvette, France}
\affiliation{Centre d'Analyse et de Math\'ematique Sociales (CNRS/EHESS) 54 Avenue de Raspail, 75006 Paris, France}

\begin{abstract}

The stochastic block model is widely used to generate graphs with a community structure, but no simple alternative currently exists for hypergraphs, in which more than two nodes can be connected together through a hyperedge. We discuss here such a hypergraph generalization, based on the clustering connection probability $P_{ij}$ between nodes of communities $i$ and $j$, and that uses an explicit and modulable hyperedge formation process. We focus on the standard case where $P_{ij}=p\delta_{ij}+q(1-\delta_{ij})$ when $0\leq q\leq p$ ($\delta_{ij}$ is the Kronecker symbol). 
We propose a simple model that satisfies three criteria: it should be as simple as possible, when $p = q$ the model should be equivalent to the standard hypergraph random model, and it should use an explicit and modulable hyperedge formation process so that the model is intuitive and can easily express different real-world formation processes. We first show that for such a model the degree distribution and hyperedge size distribution can be approximated by binomial distributions with effective parameters that depend on the number of communities and $q/p$. Also, the composition of hyperedges goes for $q=0$ from `pure' hyperedges (comprising nodes belonging to the same community) to `mixed' hyperedges that comprise nodes from different communities for $q=p$. We test various formation processes and our results suggest that when they depend on the composition of the hyperedge, they tend to favor the dominant community and lead to hyperedges with a smaller diversity. In contrast, for formation processes that are independent from the hyperedge structure, we obtain hyperedges comprising a larger diversity of communities. The advantages of the model proposed here are its simplicity and flexibility that make it a good candidate for testing community-related problems, such as their detection, impact on various dynamics, and visualization.

\end{abstract}

\maketitle

\section{Introduction}

The stochastic block model (SBM) is a simple model of a graph with communities. This generative model for data benefits from a ground truth for the communities, and has
been widely used as a canonical model for community detection (see the review \cite{Abbe2017}). The standard community detection problem \cite{Fortunato} is the statistical recovery problem of the clusters: from the graph, can we reconstruct the sets of nodes
 that describe the communities? Depending on the parameter values there are different regimes such as impossible to recover, easy, etc. \cite{Mossel,Decelle,Abbe2014}. Most of these works concerned networks, but recent analysis of complex systems such as in systems biology \cite{Krieger2021}, face-to-face systems \cite{Cencetti:2020}, collaboration teams and networks \cite{Patania:2017,Juul:2022}, ecosystems \cite{Grilli:2017}, the human brain \cite{Petri:2014,Giusti:2016}, document clusters in information networks, or multicast groups in communication networks \cite{Gao2012}, etc. showed
that graphs provide a limited view, and in order to include
higher-order interactions involving groups of more than two units, it is necessary to refer to hypergraphs \cite{Battiston2020,Battiston2021,Bianconi2021}. These objects that are a natural extension of usual graphs, consist in allowing edges to connect an arbitrary
number of nodes. These `hyperedges' constructed over a
set of vertices define what is called a hypergraph \cite{Berge:1984,Bretto:2013}. More formally, a hypergraph is defined as $H = (V, E)$ where $V$ is a set of elements (the vertices or nodes) and $E$ is the set of hyperedges where each hyperedge is defined as a nonempty subset of $V$.

Given the recent importance of these objects for describing cases that are not well described by graphs, it is natural to extend existing models. For instance, the random Erdos-Renyi graph can be extended in a number of ways, such as in \cite{barthelemy2022class} (and references therein). Growing hypergraph models were also proposed in \cite{Wang:2010,bianconi2017emergent,courtney2017weighted,Krapivsky, roh2023growing}
and in particular, introduced preferential attachment in this context \cite{Wang:2010, roh2023growing}. In \cite{Krapivsky}, random recursive hypergraphs grow by adding at each time step a vertex and an edge by joining the new vertex to a randomly chosen existing edge. 

Some other generalizations were proposed, such as a null model for hypergraphs with fixed node degree and edge dimensions \cite{Chodrow2019,courtney2017weighted} which generalizes
to hypergraphs the usual configuration model \cite{Molloy1998}. Other models for higher-order interactions were described in the
review \cite{Battiston2020} such as bipartite models, exponential graph models, or motifs models, but it is fair to say that in general hypergraph modeling is less developed than its network counterpart. Most of these hypergraph models are introduced in the mathematical literature and are usually thought of as immediate generalizations of classical graph models, where it is assumed that all hyperedges have the same size (and are called $k-$uniform hypergraphs when the size of hyperedges is $k$), which is a strong constraint. 

Concerning the SBM, some papers proposed an extension to hypergraphs (see for example \cite{Chodrow:2021, Ruggeri1}). Most of the recent models such as in \cite{Ghosh,Kim,Angelini,Ke}, consider hypergraphs with fixed degree and/or edge sizes (which in general do not correspond to empirical observations).  More precisely, in \cite{Chodrow:2021}, the authors propose a generative approach to hypergraph clustering based on a degree-corrected hypergraph stochastic blockmodel. This model generates clustered hypergraphs with heterogeneous degree distributions and hyperedge sizes, based on a probability distribution over the space of possible hypergraphs. This is obviously a very general model, but difficult to use and to connect with real hypergraphs and processes at play in the formation and evolution of these structures. In \cite{Kim}, the authors study the problem of community detection in a random hypergraph model which they call the stochastic block model for $k-$uniform hypergraphs (a model introduced in \cite{Ghosh}). Each hyperedge appears in the hypergraph independently with a probability depending on the community labels of the vertices involved in the hyperedge. 
Other recent works constructed random hypergraphs models with communities and generalized the LFR community detection benchmark\cite{abcd} or the  Poisson stochastic block model\cite{Ruggeri1,Ruggeri2023}.
In particular, in \cite{Ruggeri1}, the authors proposed a sampling approach where hyperedges are generated according to some probability and where nodes belong to different communities. As the authors of \cite{Ruggeri2023} noted, even if hypergraphs are widely adopted tools to examine systems with higher-order interactions, we are still lacking a theoretical understanding of their detectability limits in the community detection problem \cite{Angelini,Ke,Zhou,Conti,Dumi}. In this respect, it seems important to develop simple models of hypergraphs with communities in order to test various detection methods, but also to explore the impact of such a structure on various dynamics \cite{Majhi} such as diffusion \cite{Carletti}, synchronization \cite{Skardal,Zhang2023} or spreading \cite{Iacopini,Chowd,Neu}.

Most of these models are however very general, difficult to code, and are usually not well adapted to the case of growing hypergraphs. Indeed, these works take, in general an `ensemble' perspective where the hypergraph is an element chosen according to some distribution in a large set of hypergraphs. Here, we choose another setting, where we perform loops over the hyperedges and the nodes and ask if a given node joins with an existing hyperedge. The crucial ingredient here is thus the formation process of hyperedges. Hyperedges are, in fact, groups of nodes and we can envision that different real-world networks correspond to different formation processes of these groups. This is particularly relevant in the case of growing structures: a new node arrives in the system and will connect to existing hyperedges, or in other words, will join an existing group of nodes. We thus propose such a simple model where we can explicitly introduce the hyperedge formation process and study its impact on the hypergraph structure. 

\section{Model}

\subsection{The stochastic block model}

We assume to have $N$ nodes and a partition of these nodes into $K$ disjoints communities (or subsets) $C_1, C_2, \ldots, C_{K}$. The (symmetric) probability matrix $P$ of connections is assumed to be known. More precisely, for constructing the SBM, two vertices $u\in C_i$ and $v\in C_j$ are connected by an edge with probability $P_{ij}$. The planted partition model is the special case given by a probability matrix of the form (in some studies, this restricted model is what is called the stochastic block model)
\begin{align}
P(p,q) = \begin{bmatrix} 
    p & q & \dots \\
    \vdots & \ddots & \\
    q &        & p 
    \end{bmatrix}
    \label{eq:Ppq}
\end{align}
Thus, two vertices within the same community share an edge with probability $p$, while two vertices in different communities share an edge with probability $q$. This simple form allows us to tune the ratio intra- versus inter-connections. When $p > q$ this model is called an assortative model, while the case $p<q$ is called disassortative. For some algorithms, recovery might be easier for assortative or disassortative block models. Also, the case $p\gg q$ corresponds to the presence of well-defined clusters while other cases such as $p\sim q$ are more fuzzy. This type of model serves as a benchmark for community detection algorithms for example. Here, we will focus on this model with $0<q<p$, which corresponds to the assortative case where in general communities have a clear meaning, but our measures could be extended to the case $q>p$. 

\subsection{A hypergraph generalization}

We propose a simple model where we can explicitly introduce the hyperedge formation process and study its impact on the hypergraph structure. More precisely, we will impose a few constraints on our hypergraph generalization of the SBM: First, it should be simple, and second, the hyperedge formation process should be explicit and easy to modify according to the process studied. If we think of the nodes as individuals, the question of the probability to connect to an existing group is related to discussions about homophily (or heterophily), and more generally this framework allows us to test various group or team formation processes. Third, the connection probability $P_{ij}$ that a node of community $i$ is connected to a node of community $j$ -- which is a basic ingredient for the SBM -- should also be at the core of the model.

\begin{figure}[!]
	\includegraphics[width=0.4\textwidth]{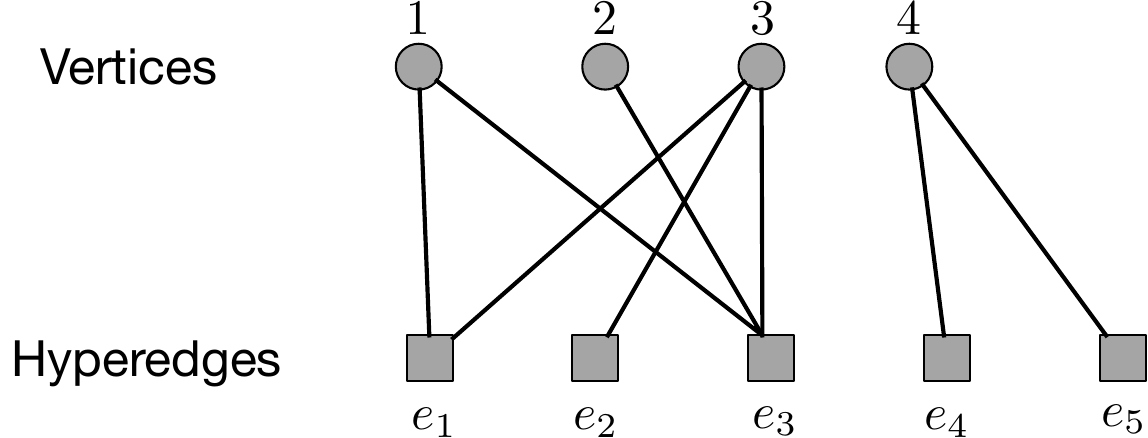}
	\caption{Bipartite representation of a hypergraph. If a node belongs to a hyperedge, it is indicated by a link in this representation.}
\label{fig:bipar}
\end{figure}

The number $N=|V|$ of vertices is called the order of the hypergraph, and the number of hyperedges $M=|E|$ is usually called the size of the hypergraph. The size $m_i=|e_i|$ of a hyperedge $e_i$ is the number of its vertices, and the degree of a vertex is then simply given by the number of hyperedges to which it is connected. A simpler hypergraph considered in many studies is obtained when all hyperedges have the same cardinality $k$ and is called a $k$-uniform hypergraph (a $2$-uniform hypergraph is then a standard graph). A simple way to represent a hypergraph is by using its bipartite description where on one side we have nodes and on the other the hyperedges. A link in this bipartite representation describes then the membership of a node in a hyperedge (see for example Fig.~\ref{fig:bipar})

We observe that in this representation, the fundamental quantity is the probability $\mathrm{Prob}(v\in e)$ that a node $v$ belongs to an edge $e$. For example, a generalization of the random Erdos-Renyi graph \cite{Erdos} can then be obtained with the following choice studied in \cite{barthelemy2022class}
\begin{align}
    \mathrm{Prob}(v\in e)=p
\end{align}
where $p\in [0,1]$.

Using this representation, we would like to find a model for a stochastic block hypergraph that satisfies a number of requirements:
\begin{itemize}
\item{} The model should be as simple as possible, where the minimal input is the number of nodes ($N$), the number of hyperedges ($E$), the set of communities ($C_1,C_2\ldots,C_K$), and the connection probability matrix $P_{ij}$ (here we will focus on the form given in Eq.~\ref{eq:Ppq}). 
\item{} When $p = q$, i.e., when the probability of a node joining a hyperedge is independent from the communities (and constant), the model should recover the standard hypergraph random model as defined in \cite{barthelemy2022class}.
\item{} The hyperedge formation process should be explicit and modulable, to express different types of hyperedge formation scenarios.
\end{itemize}

We now describe such a model. We start from $N$ vertices and a given number $E$ of hyperedges. For each hyperedge $e$, we perform a loop over the $N$ nodes and use the probability $\mathrm{Prob}(v\in e)$ to test if the node $v$ will be included in the hyperedge $e$. After a certain number of steps, a given hyperedge is of the form $e=\{v_1,v_2,\ldots,v_n\}$ where the community of $v_l$ is denoted by $C(v_l)$. When adding another node $v$, we have to define the probability $\mathrm{Prob}(v\to e)$ that $v$ will belong to the hyperedge $e$. The main point here is that this framework allows us to specify the formation process of the hyperedge, or, in other words, what drives its composition. 

There are several possible choices for this probability $\mathrm{Prob}(v\in e)$, as it could depend on the specific structure of the hyperedge $e$ and various distances between $v$ and $e$ (for example for spatial hypergraphs, the probability $\mathrm{Prob}(v\in e)$ could depend on the shortest or average distance from $v$ to $e$). At this stage, we could also integrate some knowledge about group or team formation \cite{Backstrom} and test their impact on the hypergraph structure. Here, we will focus on four simple models. 
For the first model `weighted', we choose
\begin{align}
  \mathrm{Prob}(v\to e)=\frac{1}{|e|}\sum_{u \in e}^n P_{C(u)C(v)}
  \label{eq:weighted}
\end{align}
This choice corresponds to the average of all probabilities of connections between $v$ and the nodes already in $e$. 

The model `max' is defined by the probability 
\begin{align}
  \mathrm{Prob}(v\to e)=\max_{u \in e} P_{C(u)C(v)}
    \label{eq:max}
\end{align}
In this case, the hyperedge formation process is governed by the most probable connection.  In contrast, we could imagine the case where we have some `repulsive' effect and where the presence of an individual of another group determines the connection probability. In this model `min', we then have 
\begin{align}
  \mathrm{Prob}(v\to e)=\min_{u \in e} P_{C(u)C(v)}
    \label{eq:min}
\end{align}

Finally, the last model that we will consider here depends on the most frequent community in the hyperedge: if the majority of nodes in the hyperedge $e$ are in the community $C$, the model `majority' is defined by
\begin{align}
  \mathrm{Prob}(v\to e)=P_{C(v)C}
    \label{eq:majority}
\end{align}

These strategies could reflect different real-world interaction processes. For example, a person (node) joining a discussion group (hyperedge) may reflect the `weighted' or the `frequent' strategies, where the person gauges the frequency of people he knows in the group before joining it or not. In contrast, in scenarios where nodes have some notion of risk by joining hyperedges, the `min' strategy could reflect better the behavior of the agents.

The algorithm for constructing the hypergraph with $E$ hyperedges and $N$ nodes is then as follows:
\begin{itemize}
\item{} Initialize the nodes and hyperedges: $E$ randomly chosen nodes constitute the initial hyperedges.
\item{} We perform a loop on hyperedges
\item{} For each hyperedge in this loop, we perform a loop on nodes. For each node, we test if it will join the hyperedge according to the probability for the chosen strategy (max, min, weighted, or majority).
\item{} Once the loop on hyperedges is done, we stop the code.  
\end{itemize}

\begin{figure}
  \subfloat[]{\includegraphics[width=.5\linewidth]{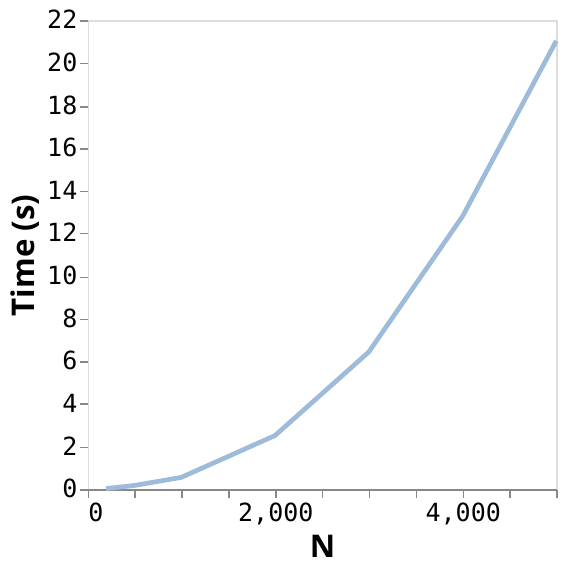}}
  \subfloat[]{\includegraphics[width=.5\linewidth]{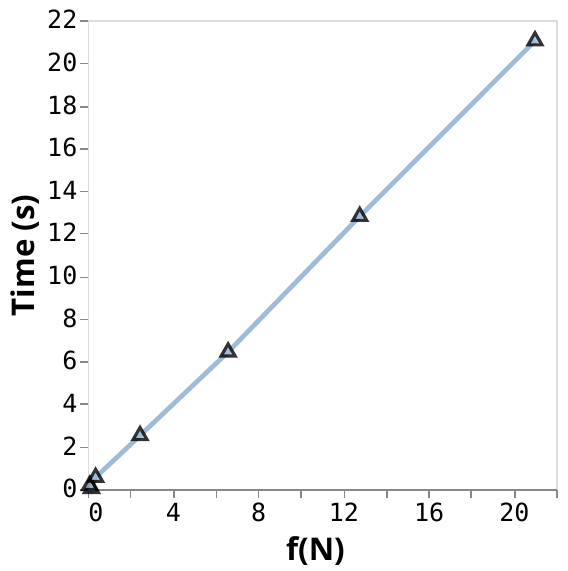}}
  \caption{Benchmark of the time (in seconds) needed to run the algorithm (a) for different values of N, with $N=E$, $K=4$, $p=\frac{100}{N}$, $q=0.4p$, using the $weighted$ strategy (each community is of the same size), and a quadratic fit (b) that shows the quadratic complexity of the algorithm ($f(N)=aN^2+bN+c$ with $a=1.01 10^{-6}$, $b=-9.8 10^{-4}$, $c=0.34$). }
      \label{fig:complexity}
\end{figure}

Here, we stop the construction of the hyperedge when all nodes are tested (in which case the probability matrix $P$ fixes the size of hyperedges), but other choices are possible. For example, we could stop the code when the size of all hyperedges reaches a value $k$ which is fixed (we then obtain what is called a $k-$uniform hypergraph), or when the size of the hyperedge reaches a random value $k$ distributed according to a distribution $P(k)$ (given empirically for example). In this paper, we focus on the simpler scenario of a fixed number of hyperedges that have a value fixed by the matrix $P$, which does not necessitate additional user-given constraints. 
In all cases, since we iterate over the nodes and hyperedges in a nested loop, the algorithm's complexity is of $O(NE)$. When $N \sim E$, this can be approximated to $O(N^2)$. We ran a small benchmark shown in \autoref{fig:complexity} that confirms the quadratic complexity of the algorithm when $N \sim E$.

\section{Simulations and results}

Many measures are available for hypergraphs: walks, paths and
centrality measures can be defined for these objects. Other measures such as the
clustering coefficient can be extended to hypergraphs
\cite{Zlatic:2009, Chodrow2019,Battiston2020,Joslyn2020,Aksoy2020}. A recent study investigated the occurrence of higher-order motifs
\cite{Lotito2022} and community detection was also considered
\cite{Turnbull2019,Ke,Chodrow:2021,Ruggeri1,Ruggeri2023,Dumi}. We can also measure the statistics of the intersection between two hyperedges as being the number of nodes they have in common \cite{Chodrow:2021}. Other measures adapted to spatial hypergraphs that take into account space can also be defined \cite{barthelemy2022class}.

Here, we will first characterize the hypergraph by its degree distribution $P(k)$, the hyperedge size distribution $P(m)$. Since our focus here is on communities we will discuss in more detail the statistics of the composition of hyperedges for different values of $q/p$. 
Most of the simulations reported are run with $E=200$, different values of $N$, and $p=0.03$, which gives hyperedges of average size $6$ for $p=q$ (we also run simulations with a higher $p=0.1$ leading to larger hyperedges, which confirm results obtained for a smaller value of $p$).

\subsection{Probability distribution for the degree and hyperedge sizes}

When $p=q$, the model reduces to a hypergraph version of the Erdos-Renyi model studied in \cite{barthelemy2022class}. In this case, the probability distributions for node degree $k$ and the hyperedge size $m$ are given by binomials
 \begin{align}
   P(k_i=k)&=\binom{E}{k}p^k(1-p)^{E-k}\\
   P(|e_i|=m)&=\binom{N}{m}p^m(1-p)^{N-m}
 \end{align}
 The average degree is then $\langle k\rangle = pE$ and the average
hyperedge size $\langle m\rangle=pN$ (which obviously satisfies the general relation $N\langle k\rangle=E\langle m\rangle$).

In the case $q\neq p$, the probability that a node connects to a hyperedge will not be constant and will depend on the hyperedge structure. When $q<p$, we introduce a smaller probability to connect, and we can expect a homogeneous behavior with probability $p$ but over a smaller effective subset of nodes of size $N^*$ and hyperedge sizes $E^*$ ($E^*$ for $P(k)$ and $N^*$ for $P(m)$). It is then reasonable to fit the degree and size distributions with binomials characterized by these effective parameters. 

\begin{figure*}[ht!]
\subfloat[]{\includegraphics[width=.49\linewidth]{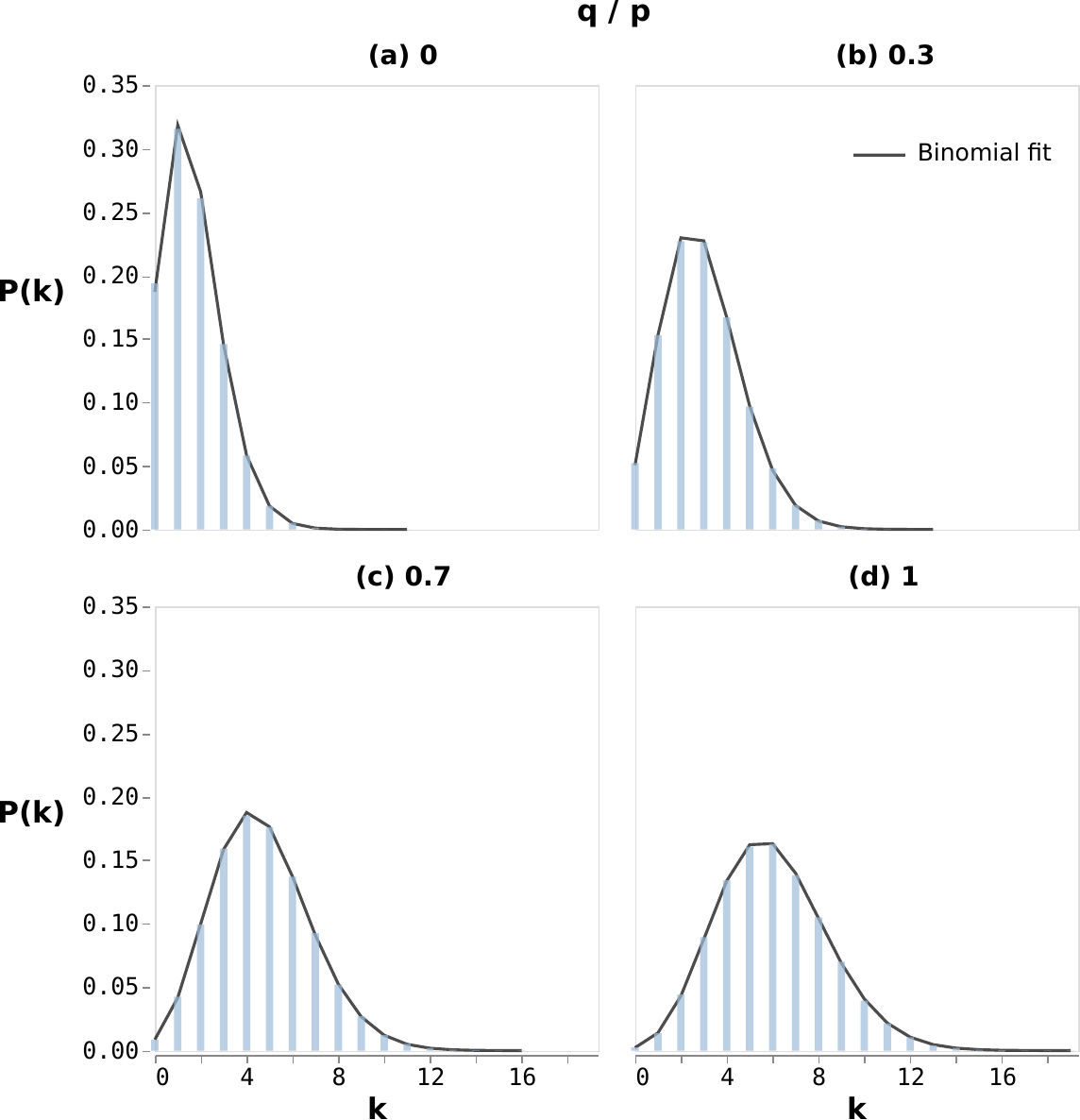}}
  \subfloat[]{\includegraphics[width=.49\linewidth]{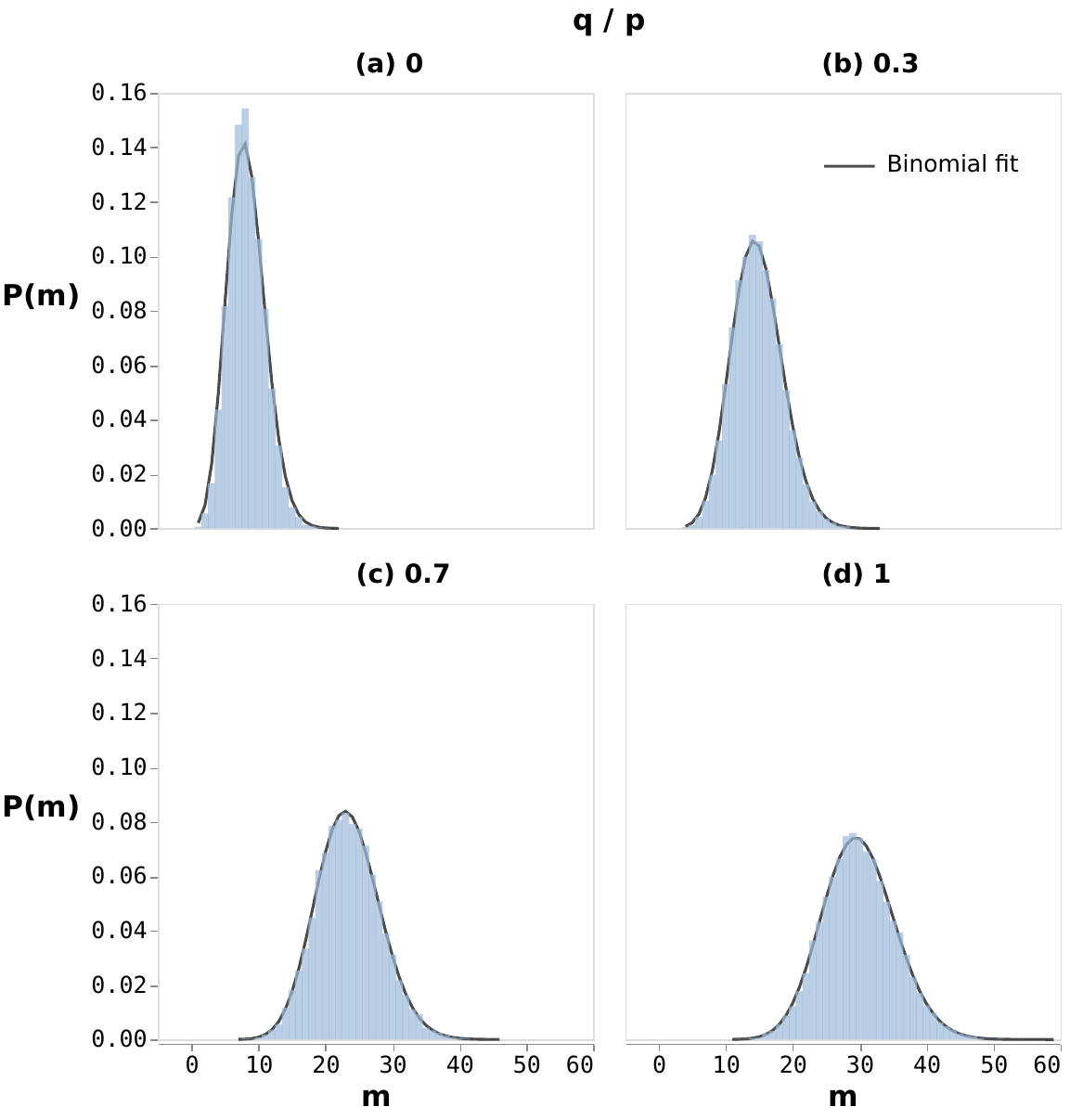}}
	\caption{Degree (a) and hyperedge size (b) distributions computed from $100$ hypergraphs with $N=1000$, $E=200$, $K=4$, $p = \frac{30}{N} = 0.03$, and $q/p = 0, 0.3, 0.7, 1$. The strategy used is the weighted probability (Eq.~\ref{eq:weighted}). The distributions are fitted with binomial distributions of parameters $(p, N^*)$.}
\label{fig:distribs}
\end{figure*}

We show for different values of $q/p$, the degree distribution in Fig.~\ref{fig:distribs} (a), and the hyperedge size distribution in Fig.~\ref{fig:distribs} (b).
This shows that even when $q\neq p$, these distributions can be approximated by binomials (which can, however, not be exact as the probability for a node to connect to a hyperedge depends on its composition and is generally not constant).

\begin{figure}[]
\includegraphics[width=0.5\textwidth]{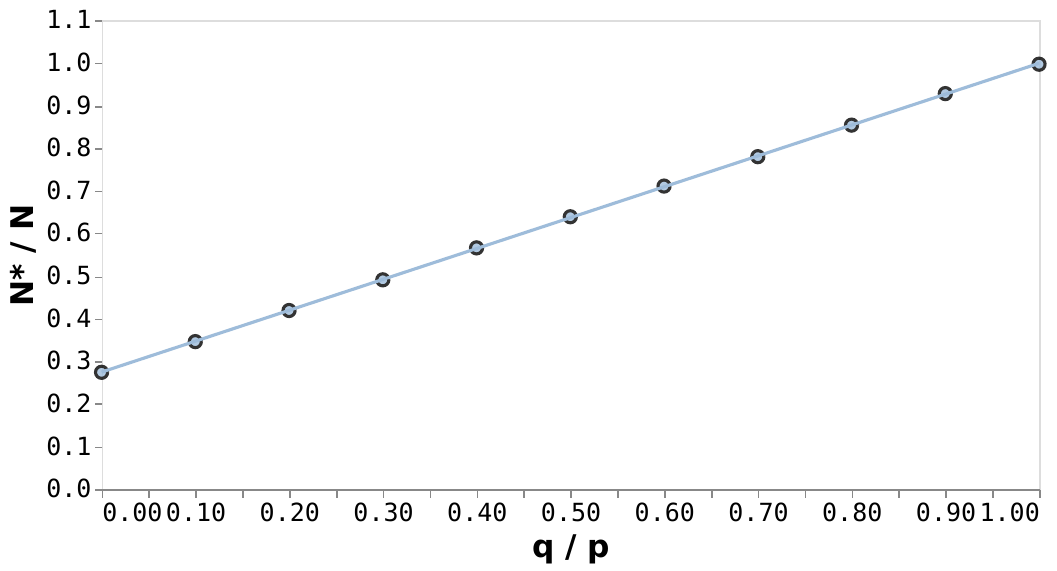}
 \includegraphics[width=0.5\textwidth]{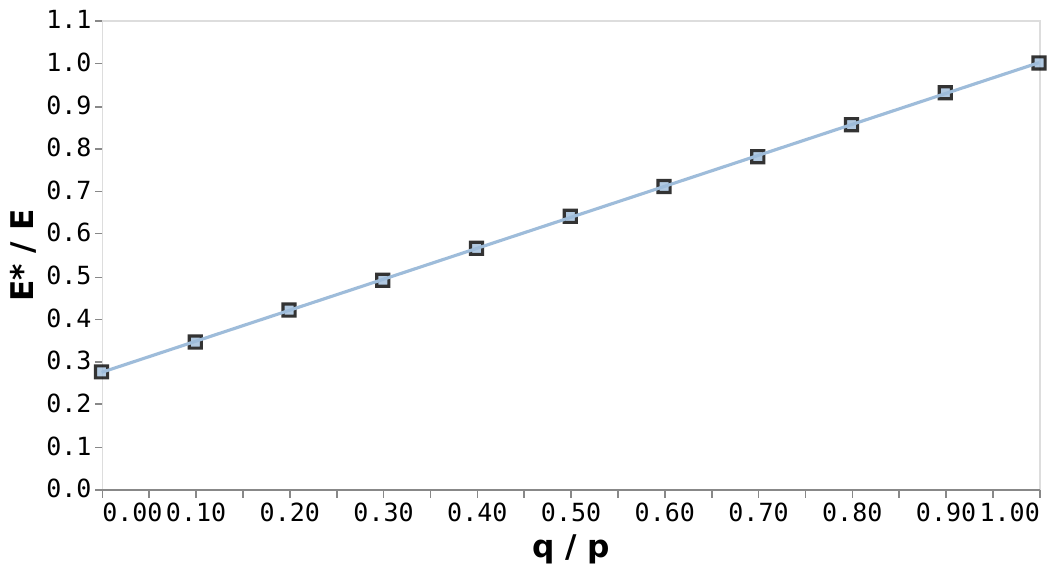}
	\caption{Effective normalized parameters (top: $N^*/N$, and bottom: $E^*/E$) obtained by fitting with binomials the degree and size distributions. The distributions are computed using $100$ hypergraphs generated with $N=1000$, $E=200$, $K=4$, $p = \frac{30}{N}=0.03$ and the weighted formation process. The lines represent linear fits of the form $a(q/p)+b$ where $a=0.73$ and $b=0.27$ for both plots, in agreement with our argument (Eq.~\protect\ref{eq:argu}).}
\label{fig:binFit}
\end{figure}

We fit these distributions with binomials with effective parameters $N^*$ and $E^*$ and plot in Fig.~\ref{fig:binFit} these parameters versus $q/p$.
We observe in this Fig.~\ref{fig:binFit}, that the effective parameters vary linearly with $q/p$. We then expect the following form
\begin{align}
  \frac{E^*}{E}=\frac{N^*}{N}=\left(1-\frac{1}{K}\right) \frac{q}{p}+\frac{1}{K}
  \label{eq:argu}
\end{align}
confirmed by our numerical experiments shown in Fig.~\ref{fig:binFit}. Indeed, for $q=0$, each community sees itself only which implies $E^*/E=N^*/N\approx 1/K$, and for $q=p$, we expect $E^*\approx E$ and $N^*\approx N$. The numerical fact that $E^*/E\approx N^*/N$ supports the consistence of this approximation.

\begin{figure}[h!]
{\includegraphics[width=1\linewidth]{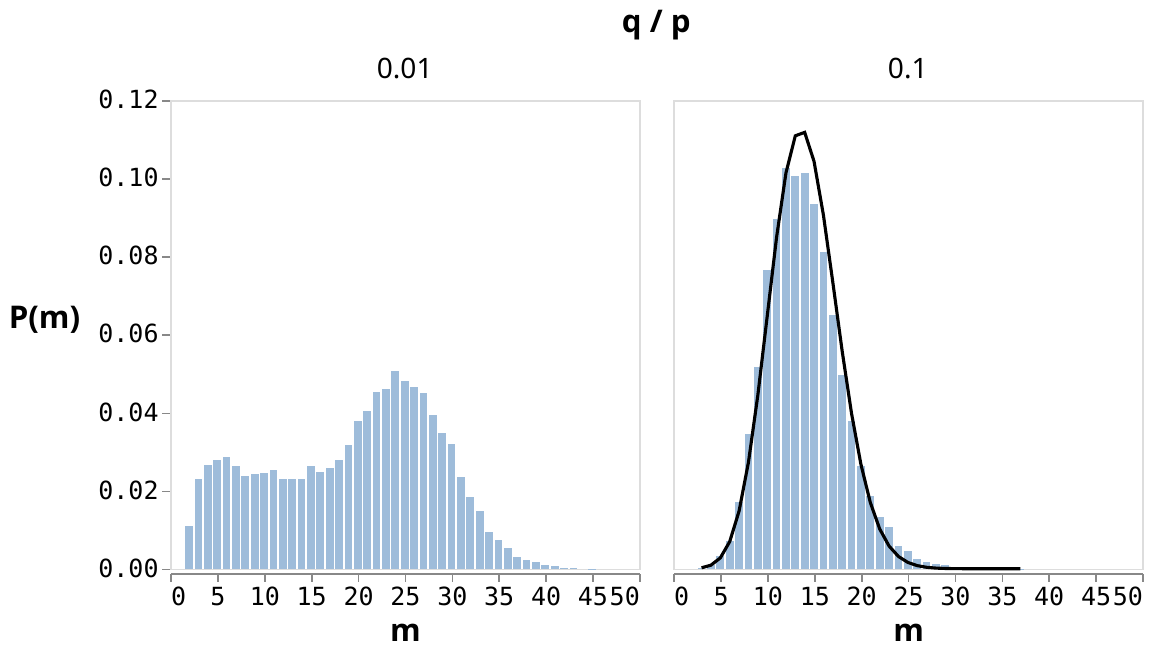}}
\caption{Hyperedge size distribution computed from $100$ hypergraphs with $N=1000$, $E=200$, $K=4$ $p = 0.1$, and $q/p = 0.1, 0.01$, with the $min$ strategy.}
\label{fig:bimodal}
\end{figure}

These results show that for this model, the degree and size distributions are therefore not very complex and can reasonably be approximated by binomials. In our simulations, the  binomial fit indeed works well for all strategies and binomial distribution were indeed observed in real world networks such as in \cite{Zhou}. Other studies (see for example \cite{Kook,Lee} and references therein) show empirical results with broad distributions for example. In order to reproduce this type of distributions, we certainly have to bring a crucial modification to the model (such as another strategy for example), but we note that the hyperedge distribution drifts away from a binomial distribution for the $min$ and $max$ strategies when $q \ll p$. In this case, the hyperedge distribution starts to take a bimodal shape, illustrated in \autoref{fig:bimodal}.
This can be explained by the following `threshold' mechanism: with the $min$ strategy, most hyperedges will be pure, as adding a node from another community is associated with the very low probability $q$. However, when this happens, the hyperedge becomes mixed, and from this moment, every node tested from any community will have a very low probability of joining the hyperedge (caused by the very low $q$). We, therefore, see very different sizes for pure and mixed hyperedges. We see a similar threshold mechanism for the $max$ strategy, but the other way around: mixed hyperedges have a higher size on average than pure hyperedges.

\subsection{Composition of hyperedges}

The most interesting part when studying hypergraphs and communities is the composition of hyperedges. Indeed, hyperedeges comprise multiple nodes that can be or not in the same community. When a hyperedge comprises nodes of the same community only, we call it `pure', or in contrast when it brings together nodes of different communities, we refer to this sort of hyperedge as `mixed'. More generally, hyperedges can contain an arbitrary collection of nodes from different communities. It is then important to understand the impact of the formation process on the composition of hyperedges. In this section, we discuss how this composition varies for various values of $q/p$ and for different strategies. 

To characterize the composition of a hyperedge, we denote by $n_i$ the number of nodes from community $i$ (where $i=1,2,\dots,K$) in the hyperedge, and a natural quantity that describes this distribution is the Gini coefficient defined as \cite{Dixon:1987}
\begin{align}
    G_0=\frac{1}{2K^2\overline{n}}\sum_{i,j}|n_i-n_j|
\end{align}
where $\overline{n}=1/K\sum_i n_i$. The maximum value of $G_0$ is $1-1/K$ and in order to obtain a quantity in $[0,1]$, we will use the normalized value $G=G_0/(1-1/K)$. The Gini coefficient averaged over all hyperedges -denoted by $\overline{G}$ - is a  measure of the composition of hyperedges. For $\overline{G}$ small, most communities enter the hyperedges, i.e., $n_i\sim m/K$ for all $i$ (where $m$ is the size of the hyperedge). In contrast for $\overline{G}$ close to one, a few communities dominate the composition of hyperedges.

\begin{figure*}
  \subfloat[$q=0$]{\includegraphics[width=.45\linewidth]{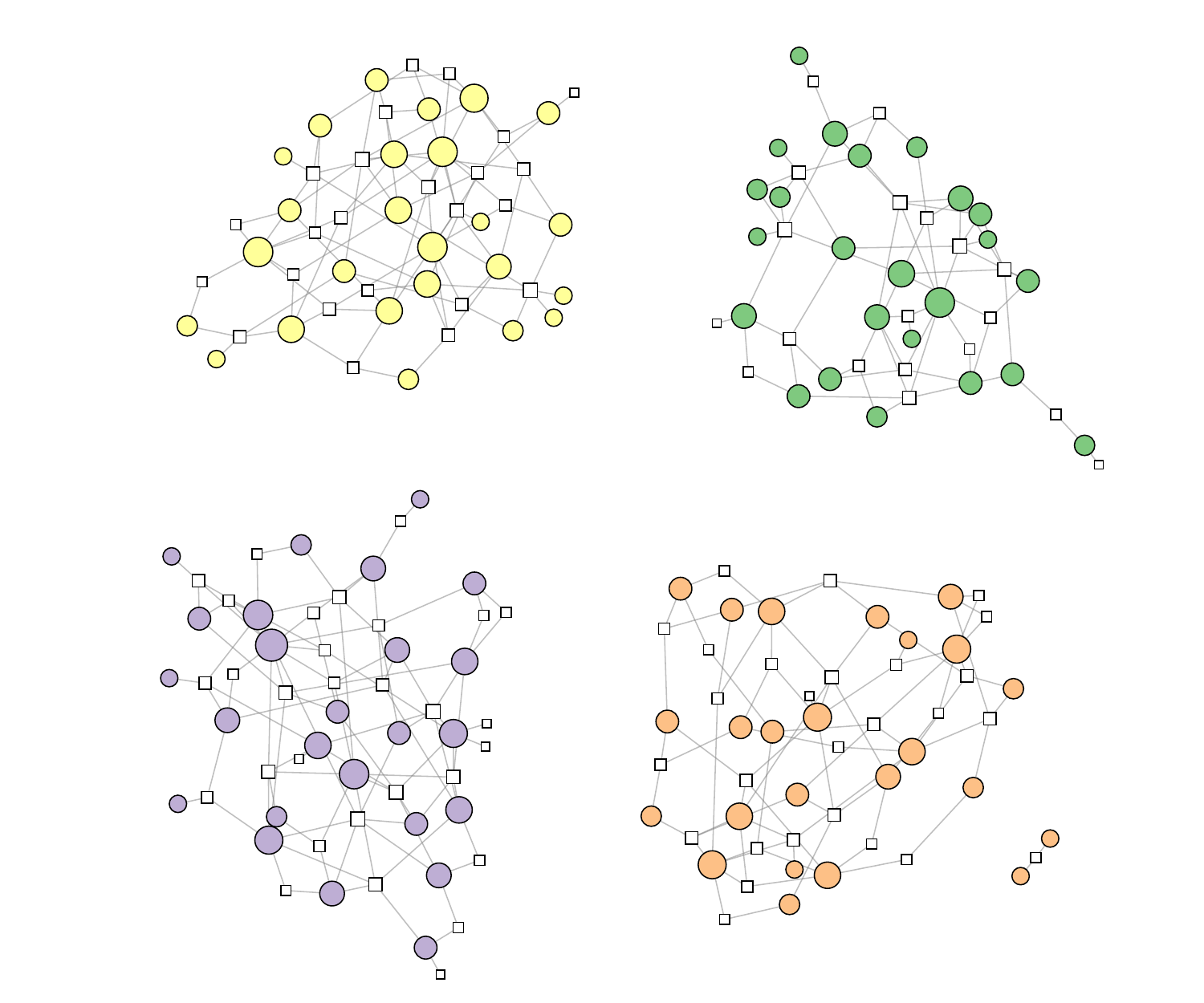}}
  \subfloat[$q=0.01$]{\includegraphics[width=.45\linewidth]{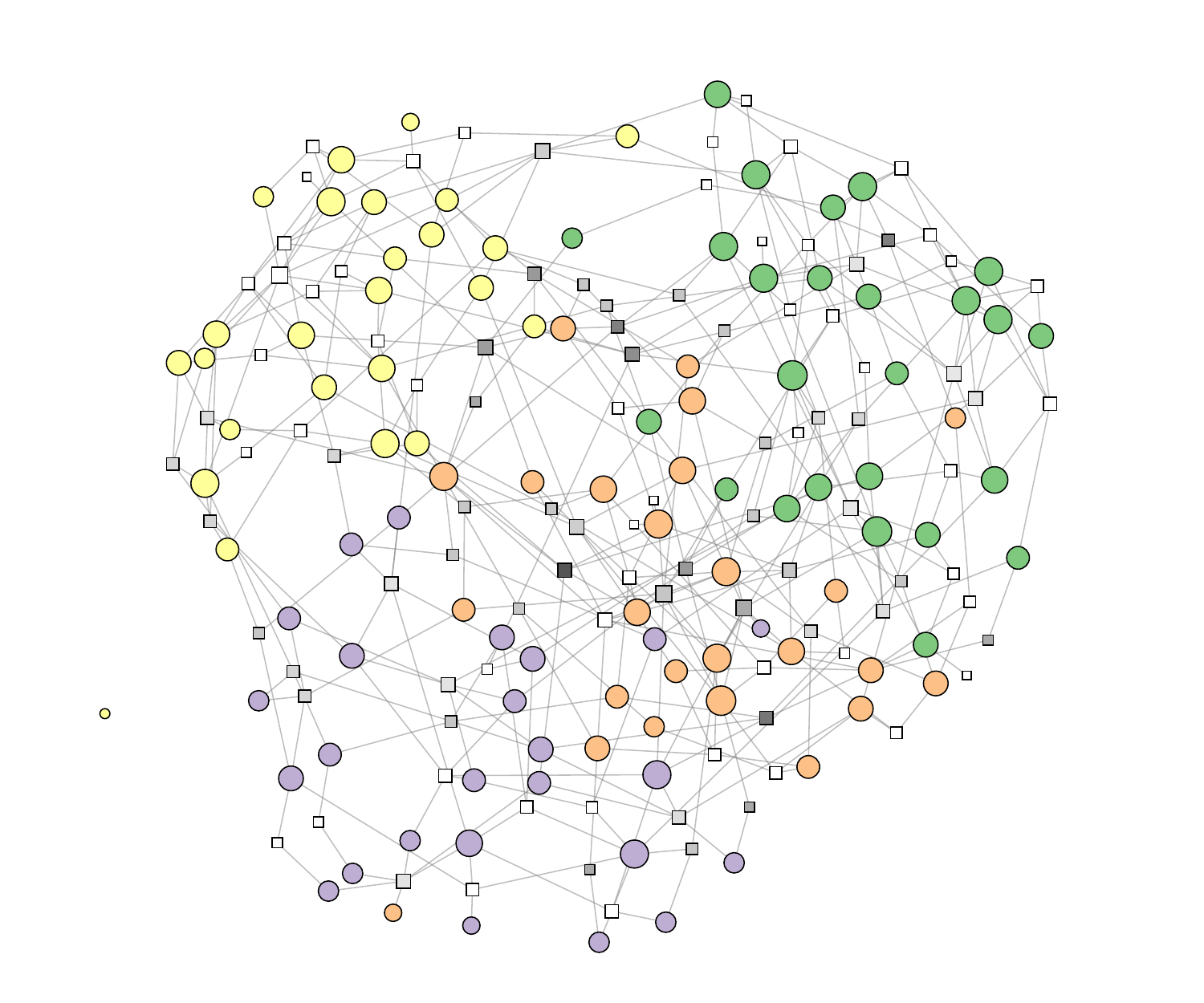}}
  
  \subfloat[$q=0.05$]{\includegraphics[width=.45\linewidth]{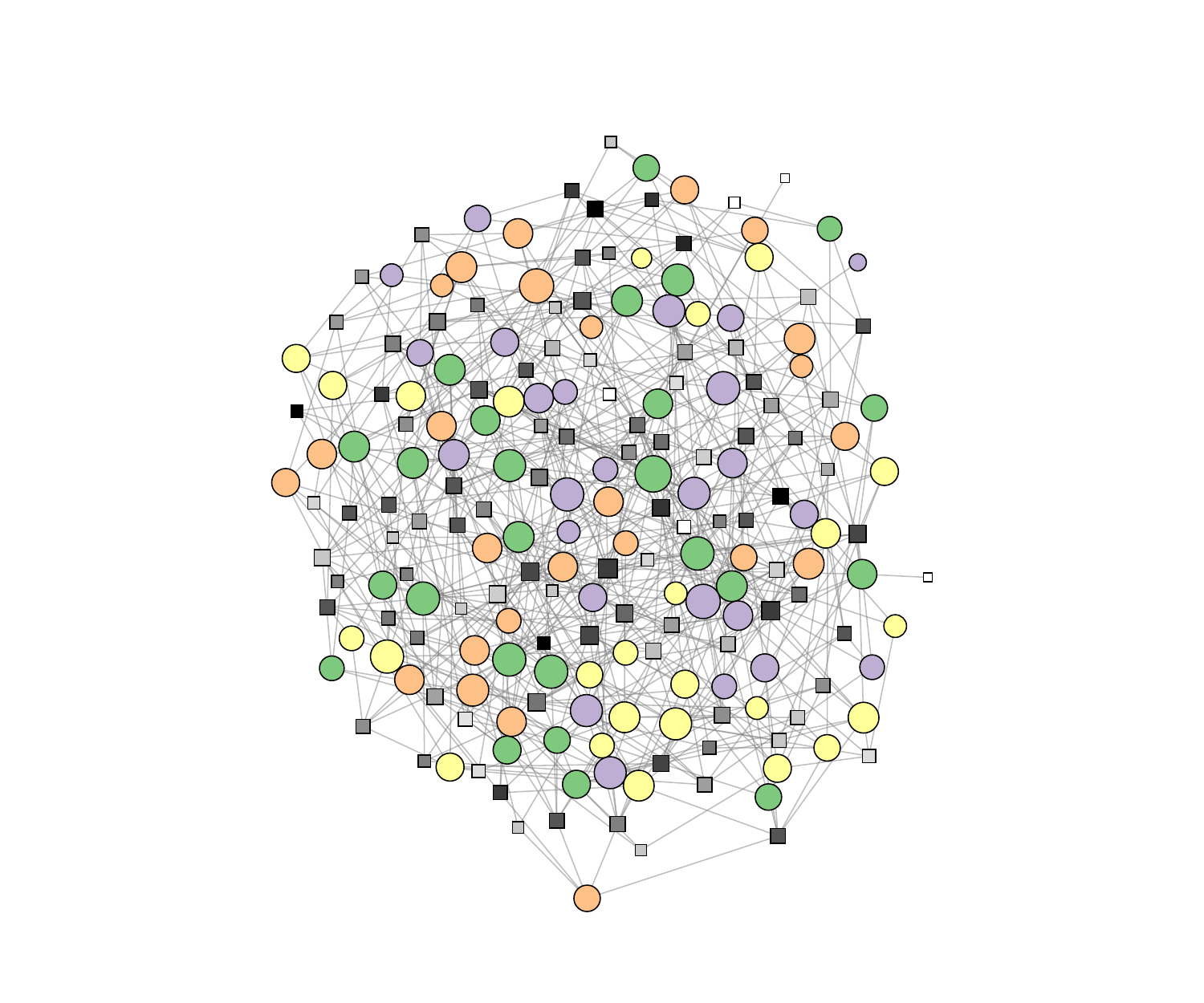}}
  \subfloat[$q=0.1$]{\includegraphics[width=.45\linewidth]{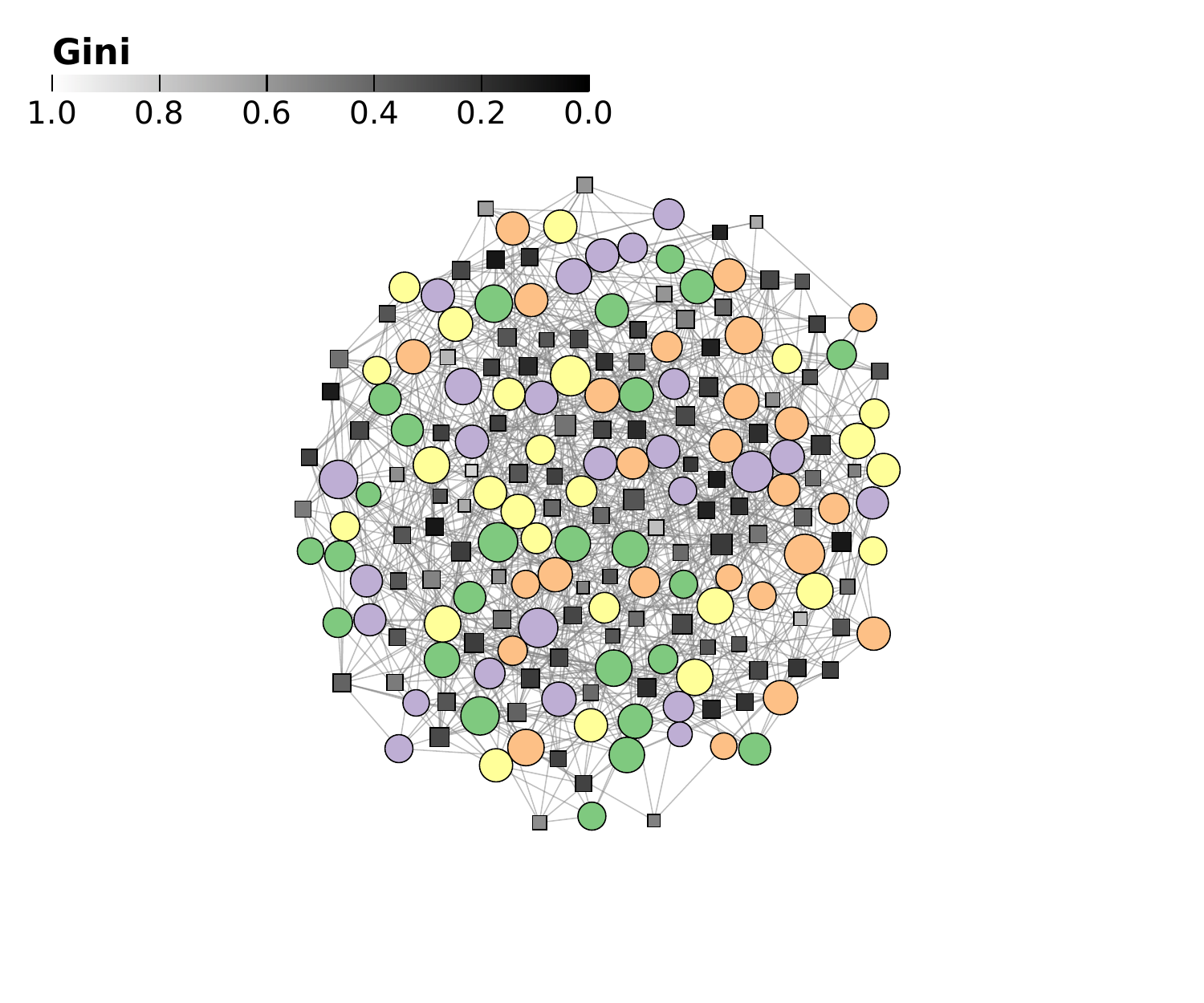}}
    \caption{Drawing of four generated hypergraphs with $N=E=80$, $K=4$, $p=0.1$, for the $majority$ strategy, and for (a) $q=0$, (b) $q=0.01$, (c) $q=0.05$, and (d) $q=0.1$. Each community is of the same size with $N/K = 20$ nodes. Hyperedges are drawn using a bipartite representation such as in \autoref{fig:bipar} using a force-directed layout implemented in D3 \cite{d3}, with circles encoding nodes and squares hyperedges. The size of nodes and hyperedges encode their respective degree and size. The color of nodes encodes their community (each color is one community), while the color of hyperedges represents their normalized Gini ($G$) values.}
    \label{fig:layouts}
\end{figure*}

The influence of $q$ and $p$ on the composition of the hyperedges is illustrated on Fig.~\ref{fig:layouts}, where we see that for low values of $q/p$ , the hyperedges tend to be pure, which reveal the community structure of the hypergraph. Interestingly, we note here the existence of hyperedges at the interface of the different communities and that are characterized by a smaller value of their Gini coefficient. For larger values, such as $q/p = 0.5$ (c), hyperedges become even more mixed, that tend to make the detection of the underlying community more complicated, if not impossible. When $p=q$ (d), the Gini coefficient $G$ is close to 0 on average, as illustrated by the darkness of the hyperedges. In other words, most hyperedges are mixed, as the process of formation of the hyperedges is community agnostic for this value.

\begin{figure*}[ht!]
        \includegraphics[width=1\textwidth]{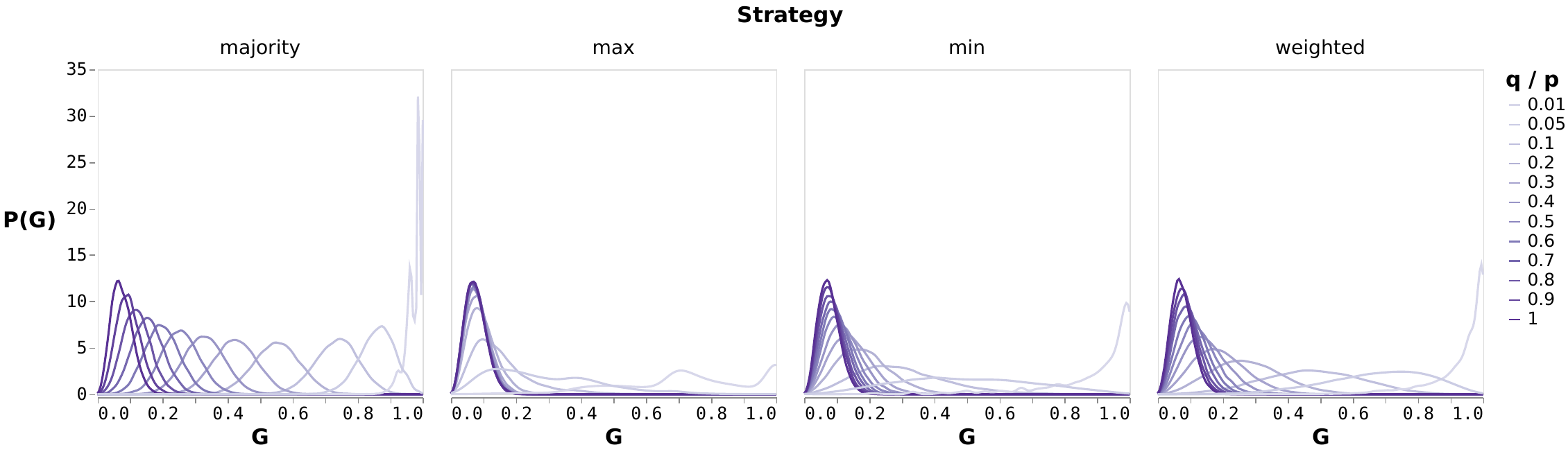}
	\caption{Distribution of $G$ for various values of $q$ ranging from 0 to $p$, with $p=0.1$, $N=2000$, $E=200$, $K=4$, and for the different formation strategies. $100$ hyperedges are generated for each combination of parameters.}
\label{fig:cDist}
\end{figure*}

More quantitatively, we show in Fig.~\ref{fig:cDist} the distribution of the Gini coefficient for different strategies and for different values of $q/p$.
When $q/p$ decreases, the probability to connect together nodes from different communities decreases and most hyperedges will be pure (or close to pure) which corresponds to $P(G)\sim \delta(G-1)$ (where $\delta$ is the Dirac delta distribution). We note that the convergence to $\delta$ is not the same for all strategies: for the 'majority' strategy, the progression of $G$ seems to follow a steady pattern from low to high values when increasing $q / p$, while the distribution of $G$ has a high variation rapidly for the other strategies. 
For $q/p$ increasing, hyperedges become more and more mixed until $q/p\sim 1$ and all communities are equally represented leading to $G\approx 0$.

\begin{figure}[ht!]
    \includegraphics[width=1\columnwidth]{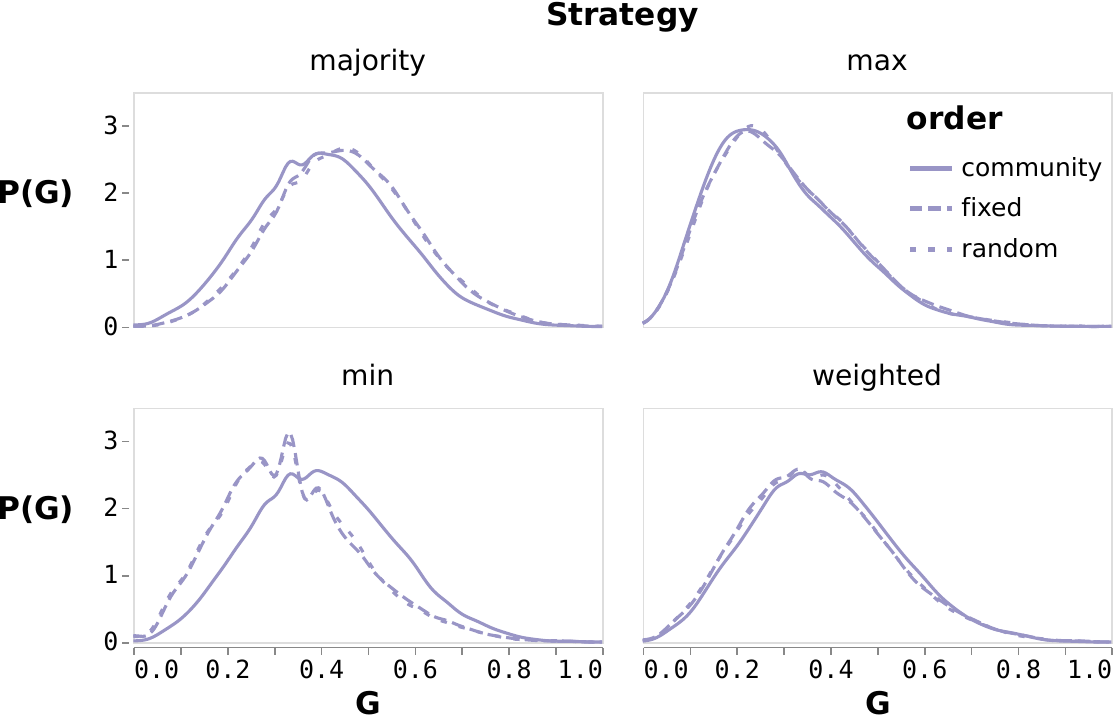}
	\caption{Distribution of $G$ for different node traversal strategies and for different hyperedge formation processes, with $p=0.03$, $q=0.4p$ $N=1000$, $E=200$, $K=4$.}
\label{fig:order}
\end{figure}

We also test how the node traversal order may affect the Gini distribution, and, therefore, the structure of the generated hypergraphs. We review three cases: a) the node ordering is random for each hyperedge, b) the node ordering is random but the same for each hyperedge (`fixed'), and c) the nodes are ordered given their community. The Gini distribution is shown in \autoref{fig:order} for the different ordering and hyperedge formation strategies. We can see that the community ordering slightly shifts the distribution to higher values, i.e., hyperedges are overall purer, for the {\it min} formation process. This is predictable, as once all nodes of community `A' are tested for a hyperedge $e$, the nodes of the following communities will have a lower probability of entering the hyperedge since several nodes of community `A' will already be in $e$.
In practice, this node traversal order does not make much sense in a modeling scenario, and we advise using the random strategy for most cases.

\begin{figure}
	\includegraphics[width=0.40\textwidth]{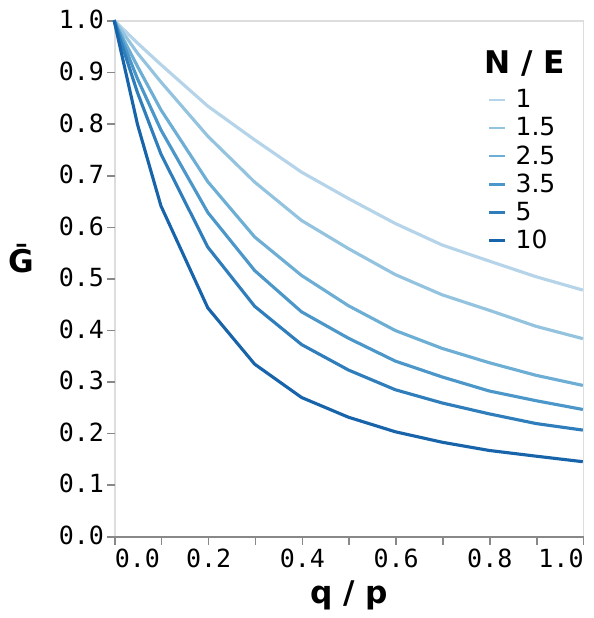}
	\caption{Average $\overline{G}$ computed over $100$ hypergraphs for $q$ ranging from $0$ to $p$, with $E = 200$, different values of $N$, and for the `weighted' strategy formation process.}
\label{fig:cMean1}
\end{figure}

\begin{figure}	\includegraphics[width=0.40\textwidth]{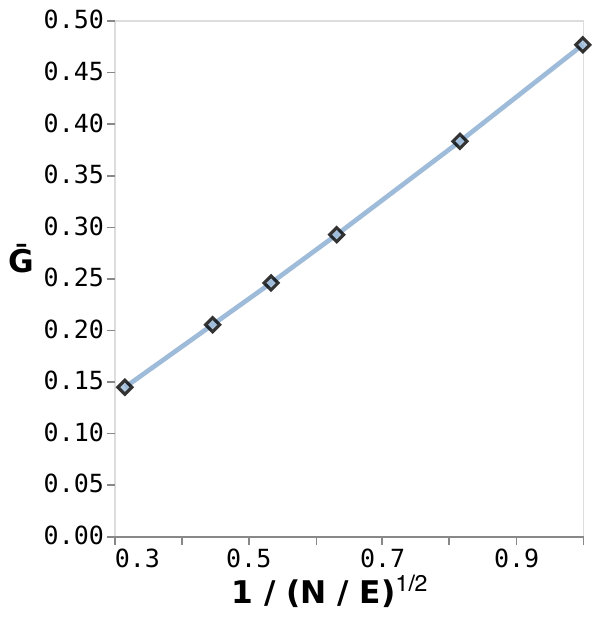}
	\caption{Average Gini coefficient $\overline{G}(q/p=1)$ versus $1/\sqrt{N/E}$ with $E = 200$, and $p=q= 0.03$. Each $\overline{G}$ is computed over $100$ generated hypergraphs.}
\label{fig:meanGvsSqrtN}
\end{figure}

In order to understand in more detail the distribution $P(G)$, we focus on the average $\overline{G}$ and the standard deviation $\sigma=\sqrt{\overline{G^2}-\overline{G}^2}$. We represent the average $\overline{G}$ versus $q/p$ for different values of $N/E$ in Fig.~\ref{fig:cMean1}.
As expected, $\overline{G}$ decreases with $q/p$, starting with pure hyperedges for $q=0$ and ending with totally mixed hyperedges for $q=p$. The value of $G(q/p=1)$ seems to depend on $N$ and we provide here a simple argument. For $q=p$, all communities are represented and if the size of the hyperedge is $m$, we have about $n_i=m/K$ nodes of community $i$. For finite sizes, we have fluctuations around this value, and typically we have
\begin{align}
  n_i=\frac{m}{K}+\frac{\xi_i}{\sqrt{N}}
\end{align}
 where $\xi_i$ are noises of order $1$. The Gini coefficient is then of order $G(q/p=1)\sim 1/\sqrt{N}$ which is confirmed by our numerical simulations in Fig.~\ref{fig:meanGvsSqrtN}.

\begin{figure}
	\includegraphics[width=0.40\textwidth]{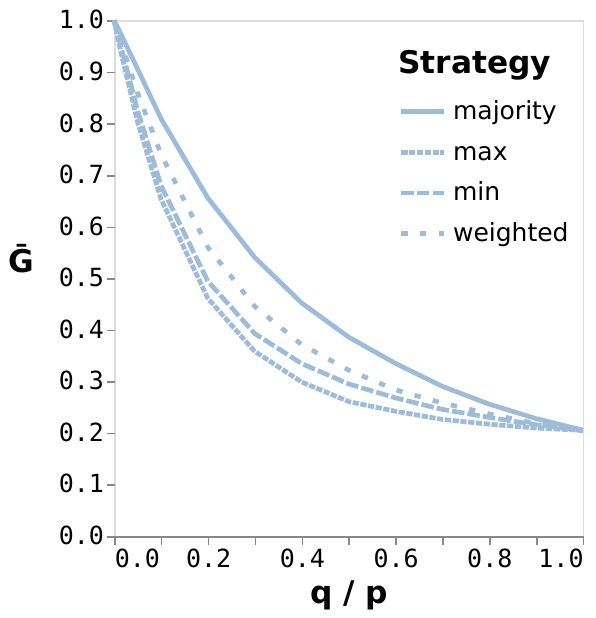}
	\caption{$\overline{G}$ (average of $G$) \rev{computed on 100 hypergraphs} for $q$ ranging from $0$ to $p$, with $E = 200$ and $N=1,000$ for different strategies.}
\label{fig:cMean2}
\end{figure}

In Fig.~\ref{fig:cMean2}, we observe that the strategy has an impact on the structure of hyperedges.
Basically, we can distinguish strategies that depend on the current fraction of the hyperedges and those that do not. The `majority' or `weighted' strategies depend on the composition of the hyperedge and tend to favor the dominant community leading to hyperedges with smaller diversity. In contrast, the `max' or `min' strategies don't depend on the fraction in the existing hyperedges and favor diversity, leading to a smaller value of the Gini coefficient. 


These arguments concern the average structure of hyperedges but say nothing about the fluctuations of their composition. In order to characterize the differences among the hyperedges of a given hypergraph realization, we will use the relative standard deviation $\Delta$ of the Gini coefficient
\begin{align}
    \Delta=\frac{\sqrt{\overline{G^2}-\overline{G}^2}}{\overline{G}}
\end{align}
A small $\Delta$ indicates that most hyperedges have typically the same value of the Gini coefficient, while a large $\Delta$ indicates a large heterogeneity of the hyperedges composition. 

\begin{figure*}[ht!]
\includegraphics[width=1\textwidth]{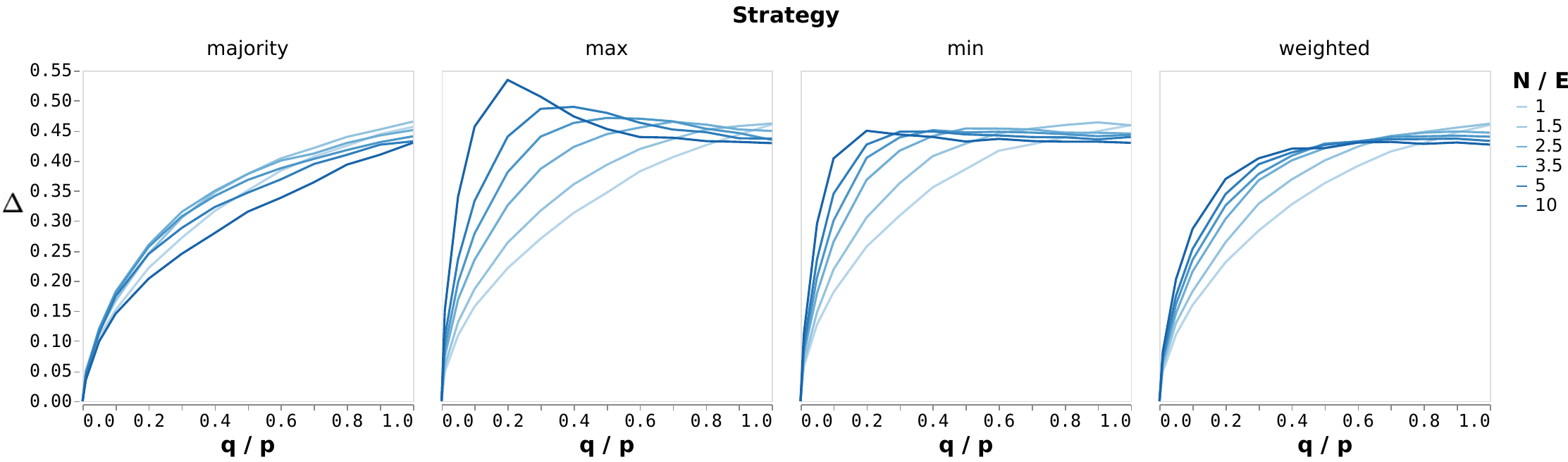}
	\caption{Normalized standard deviation $\Delta$ of $G$ for $q$ ranging from $0$ to $p$ and for different $N$. From left to right, we show the result for the different hyperedge formation strategies ($p = 0.03$, $E = 200$, $100$ generated hypergraphs for each set of parameters).}
\label{fig:cStd}
\end{figure*}

In Fig.~\ref{fig:cStd}, we show the value of $\Delta$ versus $q/p$ for different values of $N/E$ and for the different strategies.
We observe in this figure that $\Delta$ is increasing with $q/p$. For $q\approx 0$, all hyperedges are almost pure and when $q$ is increasing, fluctuations increase in general quickly. For $q=p$, all strategies lead to the same fluctuation. We observe here two remarkable facts. First, for all strategies, except for the `max' one, $\Delta$ is a monotonously increasing function of $q/p$. Indeed, for the `max' strategy, there is an optimum here around $q/p\approx 0.2$. 
\rev{
At this point, the diversity in hyperedge composition is the largest: there are a significant number of hyperedges tending toward a pure composition as $q$ is low, but hyperedges are mixing themselves more rapidly than with other strategies, thus creating a state where pure and mixed hyperedges coexist more than other strategies, explaining the optimum of $\Delta$.}
This example shows that the largest heterogeneity is not necessarily obtained for $q=p$ but depends strongly on the hyperedge formation process.

\section{Discussion}

We proposed here a simple generalization of the stochastic block model for hypergraphs. Other models were proposed recently, but here we focus on the hyperedge formation process. When a new node is considered, it connects to existing hyperedges according to some probability. This probability depends on the structure of the hyperedge and reflects the formation process of a group of nodes. This is reminiscent of group formation in various systems: when a group exists, what is the probability that a new individual will join the group? In the framework proposed here, such a process can be explicitly integrated into the model. In the simple cases we considered, our results suggest that the degree and size distributions are approximately binomials with effective parameters that depend on the connection probability (in our case, they depend on $q/p$). We also found that the strategy has indeed an impact on the composition of hyperedges, and our results suggest that the dependence on the composition is a critical parameter in the evolution of hypergraphs. Also, the strategy has a strong impact on the heterogeneity among hyperedges and we even observe that the maximal heterogeneity could be obtained for an intermediate value of $q/p$ (and not necessarily for $q=p$). Our results also suggest that when the strategy depends on the composition of the hyperedge, the resulting hypergraph has hyperedges with a small diversity (essentially made of nodes from the same community). This is in contrast when the formation process is independent from the hyperedge structure, that leads to hyperedges comprising a larger diversity of communities.

Important advantages of this model are its small number of parameters and that it is easy to implement. It can be used to model phenomena where the formation process is known or can be approximated and where we expect degree and hyperedge size distributions that follow binomial laws, such as in \cite{Zhou}. It can also be used as a null model or as a benchmark for several tasks, such as the detection and visualization of communities in hypergraphs.


{\bf Acknowledgements}
We thank Jean-Daniel Fekete for useful discussions at the beginning of this project. Much of the work was done when Alexis Pister was working independently.

{\bf Code availability}
The implementation of the model and the figures generation script are available at 
\url{https://github.com/AlexisPister/HySBM}.


\bibliographystyle{plain}

\end{document}